\documentclass[debug,overfull]{epl} 
\usepackage{epsfig}
\usepackage{changebar}
\usepackage{graphicx}
\usepackage{epsfig}
\usepackage{subfigure}
\usepackage{amsmath}
\usepackage{rotating}
\usepackage{amsfonts}
\usepackage{amssymb} 
\newlength {\oldtextheight}
\newlength {\oldheadsep}
\psfigdriver{dvips}

\begin{document}
\title{Aging  and intermittency  in a    p-spin model of a glass} 
\shorttitle{Aging of p-spin model} 
\author{Paolo Sibani}
\institute{Institut for Fysik og Kemi, SDU, DK5230 Odense M, Denmark}
\pacs{65.60.+a}{Thermal properties of amorphous solids and glasses} 
\pacs{05.40.-a}{Fluctuation phenomena, random processes, noise, and Brownian 
motion}
\pacs{61.43.Fs }{Glasses}
\date{\today}
\maketitle
\begin{abstract} 
We numerically analyze the statistics of the heat flow between 
 an aging system and its thermal bath, following a method 
  proposed and tested for a spin-glass
model  in a recent Letter (P. Sibani and H.J. Jensen, Europhys. Lett.69, 563 (2005)).
The present  system, which    lacks quenched randomness,   consists of Ising  
spins  located  on a cubic lattice,  with each plaquette
contributing  to the total energy  the  product of 
the four spins located at its corners. 
Similarly to our previous findings, 
  energy leaves the system in rare but large,  
so called  intermittent, bursts 
which are embedded in reversible and equilibrium-like   
fluctuations of zero average. 
The intermittent bursts, or  quakes,   dissipate 
the excess energy trapped in the
initial state at a rate which falls off with the inverse of the
age.  This strongly heterogeneous dynamical picture is  explained using  
the idea that quakes are triggered by energy fluctuations of record
size,  which occur  independently  within a number of thermalized domains.
From the temperature dependence of the width of the
reversible heat fluctuations we surmise  that  these domains have
an exponential density of states. Finally, we show 
that  the   heat flow consists of 
  a  temperature  independent term  and  
a term with an Arrhenius temperature dependence.      
Microscopic  dynamical and structural information can thus 
be extracted from numerical  intermittency data. This type of  analysis 
seems now within the reach  
of  time resolved micro-calorimetry techniques. 
\end{abstract} 

\section{Introduction} 
Following the  change of an external parameter, 
e.g. \ a temperature quench,  complex  
systems  unable to  re-equilibrate within 
observational  time scales undergo a  so-called  aging process.
During aging,  physical quantities slowly change as a function of the
time elapsed since the quench, a time conventionally denoted by the
term   `age'.  
In   \emph{mesoscopic}  sized   systems~\cite{Bissig03,Buisson03}
 aging   manifests itself through   a sequence of  large, so-called  
intermittent,   configurational re-arrangements. These   
events  generate     non-Gaussian tails 
 in the probability density function (PDF) of  configurational  probes 
 such as  colloidal particle displacement~\cite{Kegel00,Weeks00} and
 correlation~\cite{Cipelletti03a} 
 or voltage noise fluctuations in glasses~\cite{Buisson03a}.
The   intermittent signal   contains valuable dynamical information
which is  straightforwardly collected   in numerical  simulations
 and  which  now  seems  within  reach  of  highly sensitive
  calorimetric techniques~\cite{Fon05}. 

The present work  confirms the expected wider applicability 
of the method previously introduced in a Letter~\cite{Sibani05},
and extends the analysis to include the temperature dependence of the 
rate of energy flow. Secondly, 
it features  a more extensive  discussion of  
record dynamics~\cite{Sibani03} as a paradigm 
for non-equilibrium  aging and intermittency. 

We consider the   plaquette model~\cite{Lipowski00,Swift00}, which
belongs to a class of Ising   models   
with multiple spin interactions~\cite{Lipowski00,Swift00,Garrahan00,Davison01,Mezard02},
known to   possess central  
features of glassiness, e.g. a metastable super-cooled phase 
as well as  an aging phase~\cite{Lipowski00,Swift00}. 

In the sequel, a  short summary is given of the relevant 
properties of the p-spin model and of the
Waiting Time Algorithm~\cite{Dall01} (WTM) used in the simulations.
As a check of this   algorithm, we   calculate the known 
 equilibrium and metastability
 properties  of the model before   turning to 
 the    heat flow analysis. We show that  irreversible 
   intermittent events, so called \emph{quakes},  
 can be disentangled  from  the reversible and
equilibrium-like   fluctuations of zero average which
occur around  metastable configurations.  The age  and temperature dependences 
of the statistics of both types of events are described in detail.  
In the final section, our   findings are summarized and placed in a broader context. 
\section{The  model}
\label{model} 
We consider a set of   $N=16^3$   Ising spins, $\sigma_i = \pm 1$, placed   on a cubic
 lattice with periodic boundary conditions. The  spins 
  interact via the  `plaquette' Hamiltonian
\begin{equation}
{\cal H} = -\sum_{{\cal P}_{ijkl}} \sigma_i\sigma_j \sigma_k \sigma_l,
\label{hamilt}
\end{equation}
where  the    sum  runs   over  all the  elementary  plaquettes of the lattice,  
each  contributing the  product of the  four  spins   located at its  corners. 

By inspection, a `crystalline' ground 
state of the model is   the fully polarized state  with energy 
per spin $\epsilon_G= -3$. Other ground states can be 
obtained by  inverting  the   spins  in  any  plane orthogonal 
to  the $x$, $y$ or $z$ axes. These  symmetries lead    to a    
degeneracy $\approx 2^{3L}$, where $L$ is the linear lattice size. 

The simulations of Lipowski and Johnston~\cite{Lipowski00} show  a disordered 
equilibrium state above  $T \approx 3.6$, and, between $T=3.6$ 
and $T \approx 3.4$, 
the presence of a metastable  `super-cooled liquid'  phase. According to  
Swift et al.~\cite{Swift00}    the  energy autocorrelation function  decays in this regime  as a stretched
exponential, with no age dependence. The duration of the 
metastable plateau  diverges  as $T \downarrow 3.4$. Below 
this   temperature  aging sets in.  
\section{The WTM algorithm}
\label{WTM}
Monte Carlo algorithms   reproduce  central  
 features of aging  dynamics, 
and their use to study aging systems 
is  thus empirically, if not rigorously,  justified.    
As  in ref.~\cite{Sibani05}, the present simulations rely on  the  WTM~\cite{Dall01},  
 a rejectionless (event driven)  MC algorithm~\cite{Newman99d}, 
which  obeys detailed balance. Unlike the Metropolis algorithm, the WTM  relies on  
 an intrinsic or `global'  time variable, which is 
 akin to the clock time of  a physical system, modulo
 an overall scale factor of   little significance 
 for aging behavior, where most   quantities of interest depend on
  time ratios. 
  
   Let $b_i$ denote the energy change associated with the
   flipping of spin $i$ with fixed neighbors.  For a given initial configuration, 
   random waiting times $w_i$ are  drawn from an  exponential distribution  with average 
 $\tau_i =\exp(b_i/2T)$. 
A  WTM move identifies   the shortest waiting time,
 adds it to the  `global'  time variable,
and  flips the corresponding spin. 
Spins whose $b$ values  are affected  by the current  
move have their waiting times  recalculated.
The others do not, since the statistical properties of their 
 (memoryless) waiting process would be unchanged by a renewed evaluation.     
In    unstable  configurations,   where negative $b_i$ values  are present, 
 the WTM generates     short  waiting times, leading to    
 the nucleation of flip avalanches, which can be followed in detail.  
In  Metropolis simulations, 
  the spins  considered for update are 
randomly selected, and  events 
below the shortest  time unit of  one   MC sweep cannot be resolved. 
Thus,  while all MC algorithms with  detailed balance converge, under weak conditions,    
 to the   same  (Boltzmann) equilibrium state, they   may   do so 
 in   different ways,  see e.g.\ ref.~\cite{Gotcheva05}. 
 For  our model, a  comparison of the time dependence of the
average energy   to   results previously obtained with the Metropolis algorithm 
only shows   minor differences.  
\begin{figure}[t]
\centering
\mbox{
\subfigure[Energy relaxation]{\epsfig{figure=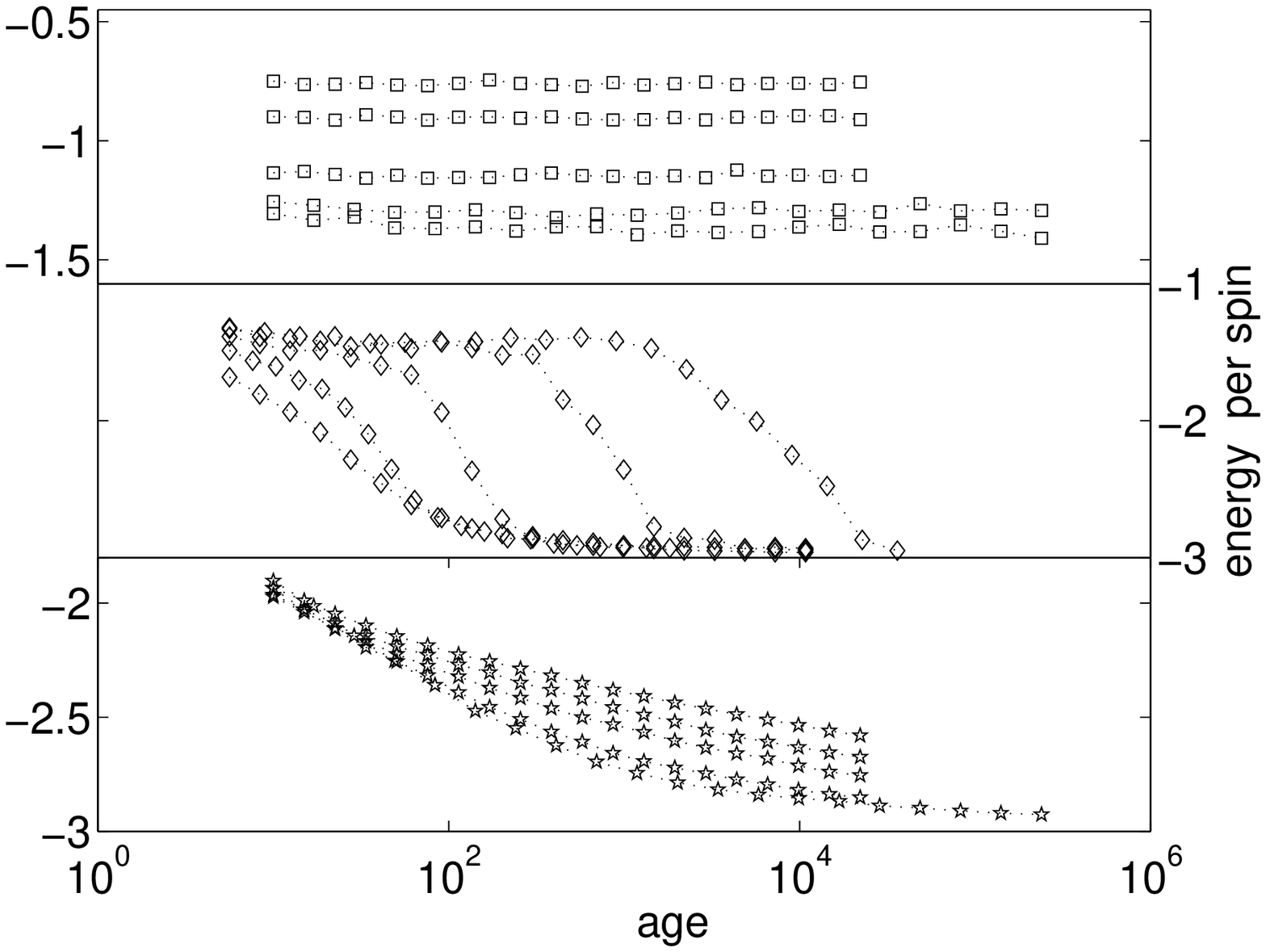,width=.47\textwidth}} \quad 
\subfigure[Heat flow PDF]{\epsfig{figure=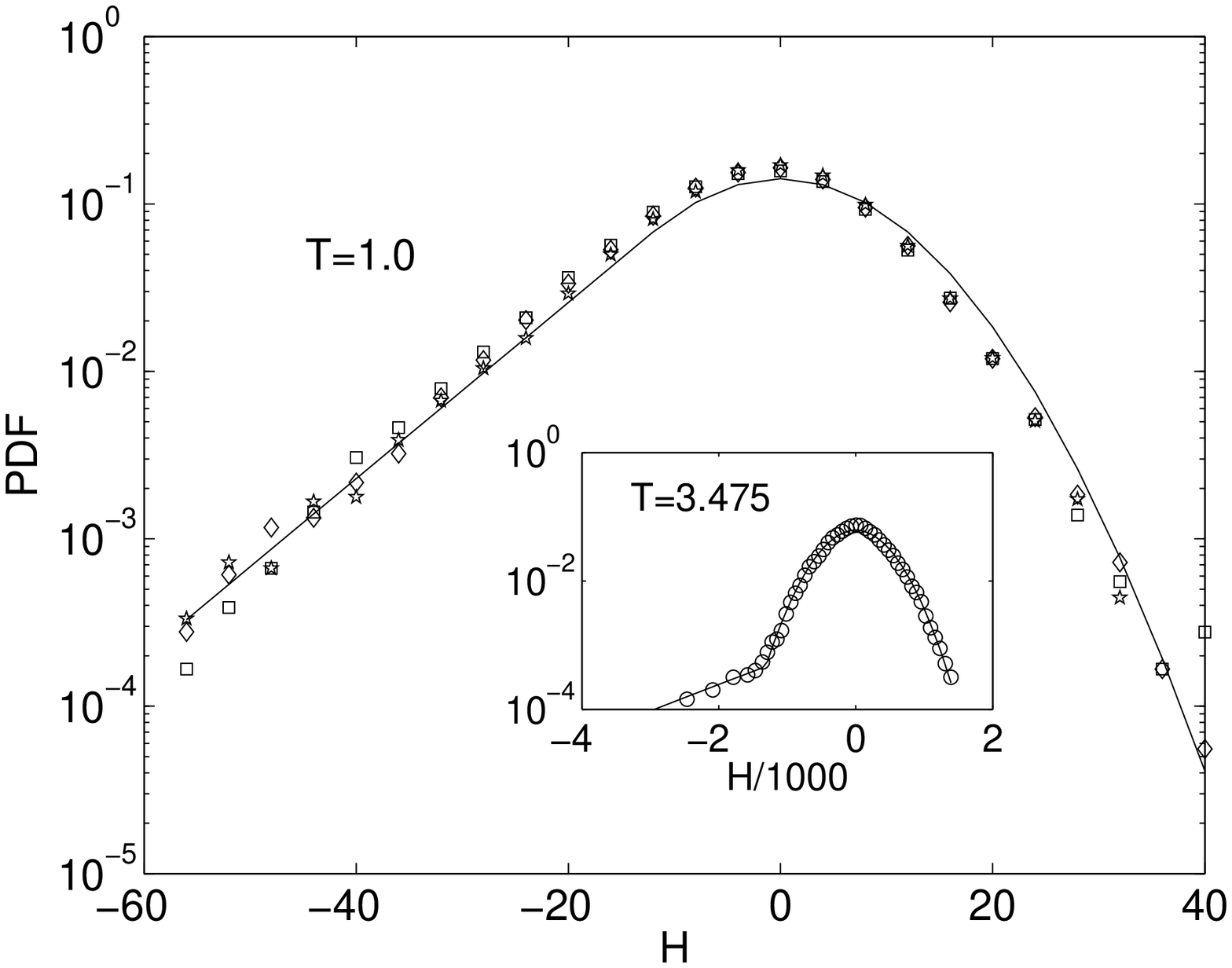,width=.47\textwidth}}} \quad 
\caption{{\em (a)}:  The energy per spin is plotted vs. the system age   for several 
temperatures, showing, from top to bottom, the presence of  thermal equilibrium,  metastability  and  aging. 
Data points are shown with symbols, the dotted lines are guides to the eye. 
In each panel, the temperatures  corresponding to  the five curves shown are, 
listed  from  top to the bottom:  $T=4.75,4.25,3.75, 3.57, 3.50$ (upper panel), $T=3.475,3.45,3.40,3.25,3.00$ 
(middle panel) and $T=1.25,1.50,1.75,2.25,2.50$ (lower panel). 
{\em (b)}: the  PDF of the heat exchanged  over an interval $\delta t$ 
at $T=1$  is  sampled in the interval  $[t_w, 3 t_w]$. The three  data sets shown have  $t_w=300, 1000$ and $5000$
respectively, with a  fixed
ratio   $\delta t/t_w =0.01$.  The  insert shows similarly collected  data for $t_w=5000$
and $T=3.475$. The line is   a fit  of the data to a function which equals 
 an exponential for sufficiently negative $H$ value and a Gaussian for the rest of the range.
 }
\label{aging_of_energy}
\end{figure}    
\section{Results}
\label{meanenergy}  
 Fig.~\ref{aging_of_energy}\emph{(a)} demonstrates 
 the three  known relaxation regimes of  this  model:
 For approximately   $T\geq 3.500$,  (top section), 
the energy  immediately  equilibrates at a high  value.
 (The  fast initial equilibration stage is omitted). 
In the interval  $3.00< T< 3.475$ (middle section),  the 
final  equilibrium  near 
the  ground state energy is reached after  dwelling in  
 a metastable  plateau. The  duration of the plateau   grows with   the
temperature, reaching its maximum near  $T=3.475$, i.e. near 
the temperature where metastability  disappears. The equilibrium 
mean energy correspondingly undergoes an abrupt change between $T=3.475$ and $T=3.5$,
in a way  reminiscent of a first order transition\cite{Lipowski00}. 
At the low temperature end of the metastable regime, the plateau gradually
softens, giving rise to a nearly  logarithmic decay of the energy. 
Approximately below $T = 2.50$ (bottom section) the energy value
reached at any stage   decreases with increasing temperature, an   extreme  
non-equilibrium situation typical for  the aging regime.  In summary, the WTM 
and the Metropolis algorithm used in other  works agree on the thermodynamical 
(average)  behavior, with the hardly significant  exception of  the  duration of the 
  metastable plateau. The latter  attains its maximum value at  $T=3.40$ under the Metropolis
   update~\cite{Swift00}, and at a slightly higher temperature in the WTM case. 
 
To investigate  the  energy flow  statistics we  choose a   
time interval $\delta t$ and partition the observation
time $t$ into $[t/\delta t]$ intervals of equal duration. The heat   $H$ exchanged
over $\delta t$  arises as    the  difference between the energies of the two 
configurations at the endpoints of the interval,  
\begin{equation}
H_n = E(t_w +(n+1)\delta t) - E(t_w +n\delta t), \quad n=1,2, \ldots [t/\delta t]
\label{H_definition}
\end{equation} 
From ref.~\cite{Sibani05}, we expect   the rate of energy flow
to fall off as $1/t_w$ (pure aging),  leading to  a     
 $\delta t/t_w$ scaling of the PDF. For the present model, 
the collapse of three PDF's shown  in the main 
 plot of Fig.~\ref{aging_of_energy}, panel \emph{(b)} confirms 
this expectation.
The PDF's    are  collected at   $T=1.0$,  
i.e.  deep in the aging regime, during time  intervals
  $[t_w,3 t_w]$, with $t_w=300, 1000$ and $5000$, 
(squares, diamonds and polygons, respectively), keeping  the  ratio  
 $\delta t/t_w$ constant and equal to $0.01$.
 The insert shows data taken at   $T=3.475$,  i.e.  at the upper edge of the 
metastable region. The calculations are repeated for $200$ 
independent trajectories, giving a total of  $20000$ data points
for each PDF. 
 
The full line is a least square fit of the data
to a continuous function comprising  an  exponential tail, $f_<(H)\propto e^{H/q}$ for $H<d<0$, and a Gaussian 
part centered at zero, $f_> \propto e^{-H^2/2{\sigma^2}_{rev}} $ for $H>d$. The  position,   $d$, 
of the joining point  is  itself a fitting parameter. As the overall scale of the PDF 
 is arbitrary,   the    three parameters  of physical significance are  
 $q, \sigma_{rev}$ and $d$.   
Figure~\ref{aging_of_energy} \emph{(b)} already   
indicates that  the  exponential  intermittent tail carries the  
bulk of  the energy flow. This also   applies 
 to the    $T=3.475$  data shown in the insert, even though   
at this   temperature, which is outside   the aging regime, 
the intermittent events are  swamped  by  the  reversible Gaussian fluctuations. 
\begin{figure}[t]
\centering
\mbox{
\subfigure[Age dependence of intermittency]{\epsfig{figure=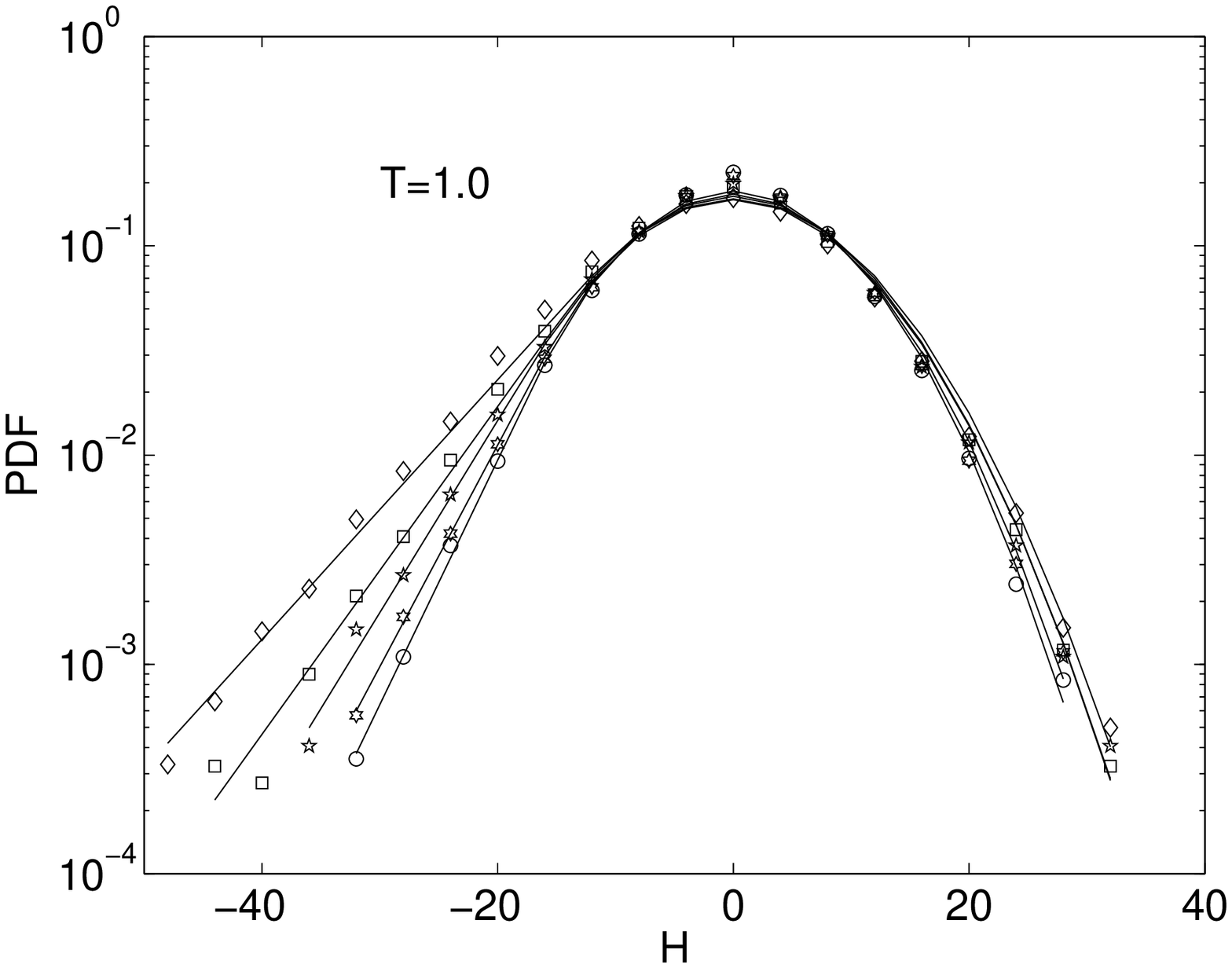,width=.45\textwidth}} \quad 
\subfigure[Rate of energy flow]{\epsfig{figure=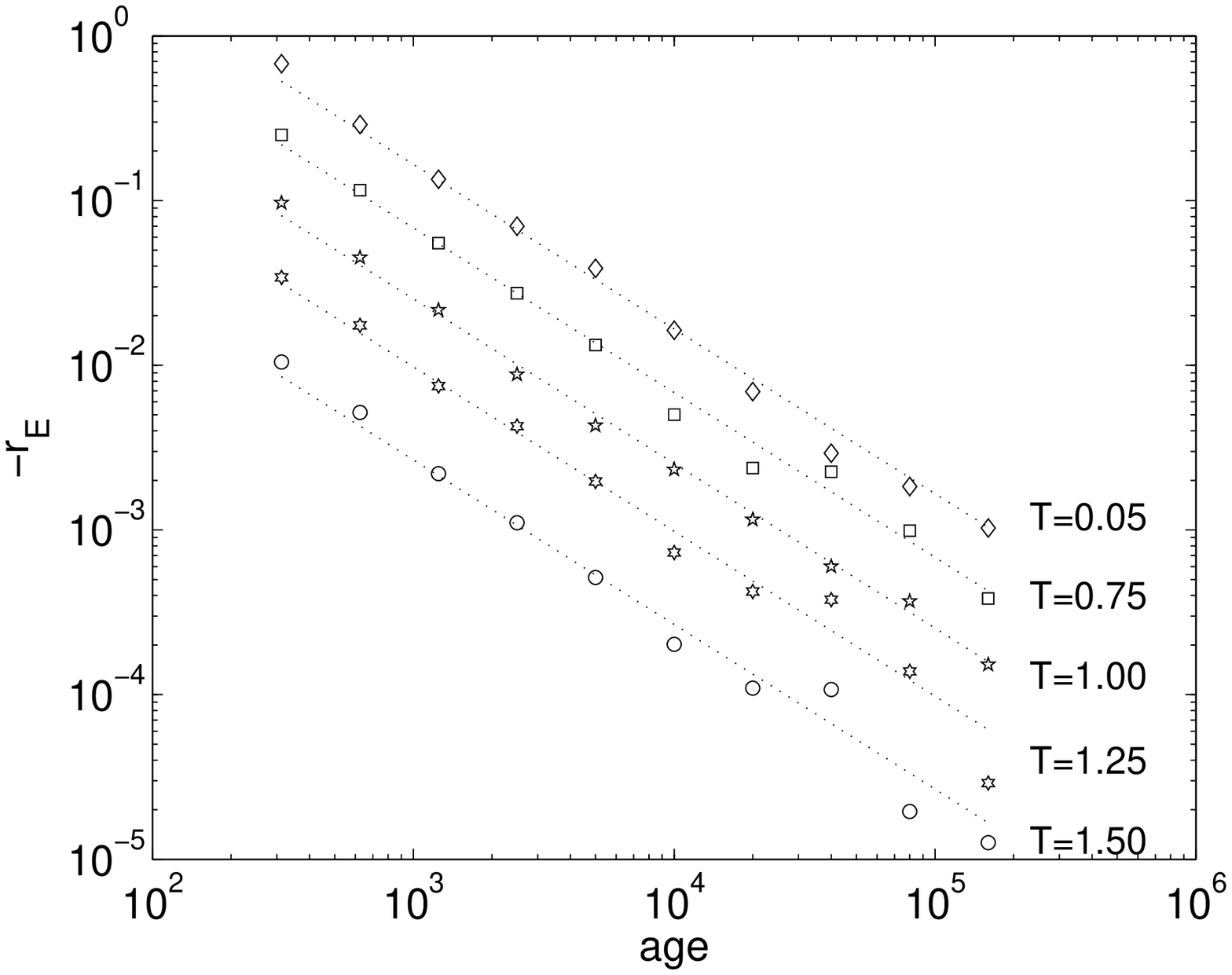,width=.45\textwidth}}} \quad 
\caption{{\em (a)}: The  PDF of the heat exchanged 
between system and thermal bath over a time $\delta t=5$. Negative values correspond to an
energy  outflow.
The data are based on $200$ independent 
runs,    taken at temperature $T=1$ in the intervals  $[t_w,t_w+500]$ for
 $t_w=1000$ (diamonds),  $t_w=2000$ (squares)  $t_w=4000$ (polygons) 
$t_w=8000$  (hexagons) and $t_w=16000$ (circles). 
{\em (b)}: The average rate of flow of the energy is plotted versus the age
for the five    temperatures shown. 
The full line has the form $y = c t_w^{-1}$, with the proportionality constant $c$ estimated as
the mean of $t_w r_E$. 
 }
\label{heat_flow}
\end{figure} 
Further  properties of the heat flow at $T=1$    are illustrated by panel \emph{(a)} of Fig. \ref{heat_flow}.
 The five  PDF's shown  are  semilogarithmic plots
of the distribution  of $80000$ values of   $H$ taken  with  $\delta t=5$.
  The values are   sampled from  $800$ independent trajectories,  
each stretching over  an `observation' interval of duration $t=500$.   
    
 For sufficiently negative $H$ values, all the  PDFs   have an  exponential  intermittent tail, which turns into 
 a bell shaped,  nearly Gaussian distribution as $H$ increases toward and beyond zero.
 The strongest intermittency  observed in these data (top curve) is for
  $t_w=10^3$. Successively  doubling   $t_w$, up to   $t_w=1.6 \times 10^4$,  
 leads to  PDF's  approaching   a  Gaussian shape. This expresses the well known 
 fact that aging systems appear equilibrated, when  observed over a time shorter 
 than  their  age. 
 Importantly, the Gaussian part of the PDF has no conspicuous age dependence. 
 
 The full lines are, as mentioned,  obtained as least square fits of the PDF data
 to a continuous, piecewise differentiable curve. 
 The parameter $\sigma^2_{rev}$   estimates the  variance of the reversible fluctuations 
 and should not be confused with the full variance of $H$, which  is  mainly determined  by the  
 tail of the PDF. 

The  instantaneous  rate of energy flow, a  constant 
for time homogeneous dynamics, 
is  here inversely proportional to the age. 
Ideally, this rate is obtained  as the ensemble  average  of the ratio 
 $\frac{H(\delta t, t_w)}{\delta t}$, i.e., 
 \begin{equation}
 r_E(t_w) \stackrel{{\rm def}}{=}  \lim_{\delta t \rightarrow 0}\frac{\langle H(\delta t, t_w)\rangle}{\delta t} .
 \end{equation} 
In practice   we (mainly) use  an ensemble of  $1000$ 
independent  trajectories,
and  perform in  addition  a time average over a 
finite  observation interval $t$.  
We use   $t=30$ for $t_w \le 2500$,  $t=100$ for $t_w \le 40000$ and
$t=500$ for $t_w > 40000$. By varying  $\delta t$   over the  interval 
 $[0.1,23]$ we ascertain that  $\langle H \rangle$ 
 is  linearly dependent on $\delta t$. Finally, we estimate 
 $r_E$ as the arithmetic average of   $H(\delta t, t_w)/\delta t$,
over the approx. $5500$ data points available at each temperature.  
Panel \emph{(b)} of Fig. \ref{heat_flow} shows
in a log-log plot the negative  energy flow rate 
versus the system age,  for five selected temperatures. 
To avoid   clutter, the  data sets belonging to  each $T$ value 
 are artificially 
shifted in the vertical direction by  dividing, in order of increasing $T$,
 by  $1,3,9,27$ and $81$.  
 The    full lines of  Fig.\ref{heat_flow} are obtained by 
 fitting the data to the  form $r_E = c/t_w$, with  the chosen value of $c$ 
 minimizing   the RMS distance to the data.

According to record dynamics~\cite{Sibani03,Sibani05},  the rate of quakes,
$r_q$ obeys 
\begin{equation}
r_q(t_w) = \frac{\alpha(N)}{t_w}, 
\end{equation}
where   $\alpha$ represents 
the   number of independent domains in the system, 
which is linearly dependent on the system size $N$, but independent 
of the temperature. The rate of energy flow  depends on the
rate of quakes and on the  amount of energy given off in each quake. 
The latter could in principle depend on both temperature and age. 
However, since     panel \emph{(b)} of Fig. \ref{heat_flow} shows that    
 $r_E \propto 1/t_w$ to a good approximation, we  conclude that 
$r_E(t_w) = r_q(t_w) e(T)$,
where the    average  amount $e(T)$
of energy exchanged  in a single  quake is given by 
\begin{equation}
e(T) = \frac{r_E \cdot t_w}{\alpha},
\label{energyperquake} 
\end{equation} 
which  only depends on $T$.
\begin{figure}[t]
\centering
\mbox{
\subfigure[Reversible fluctuations]{\epsfig{figure=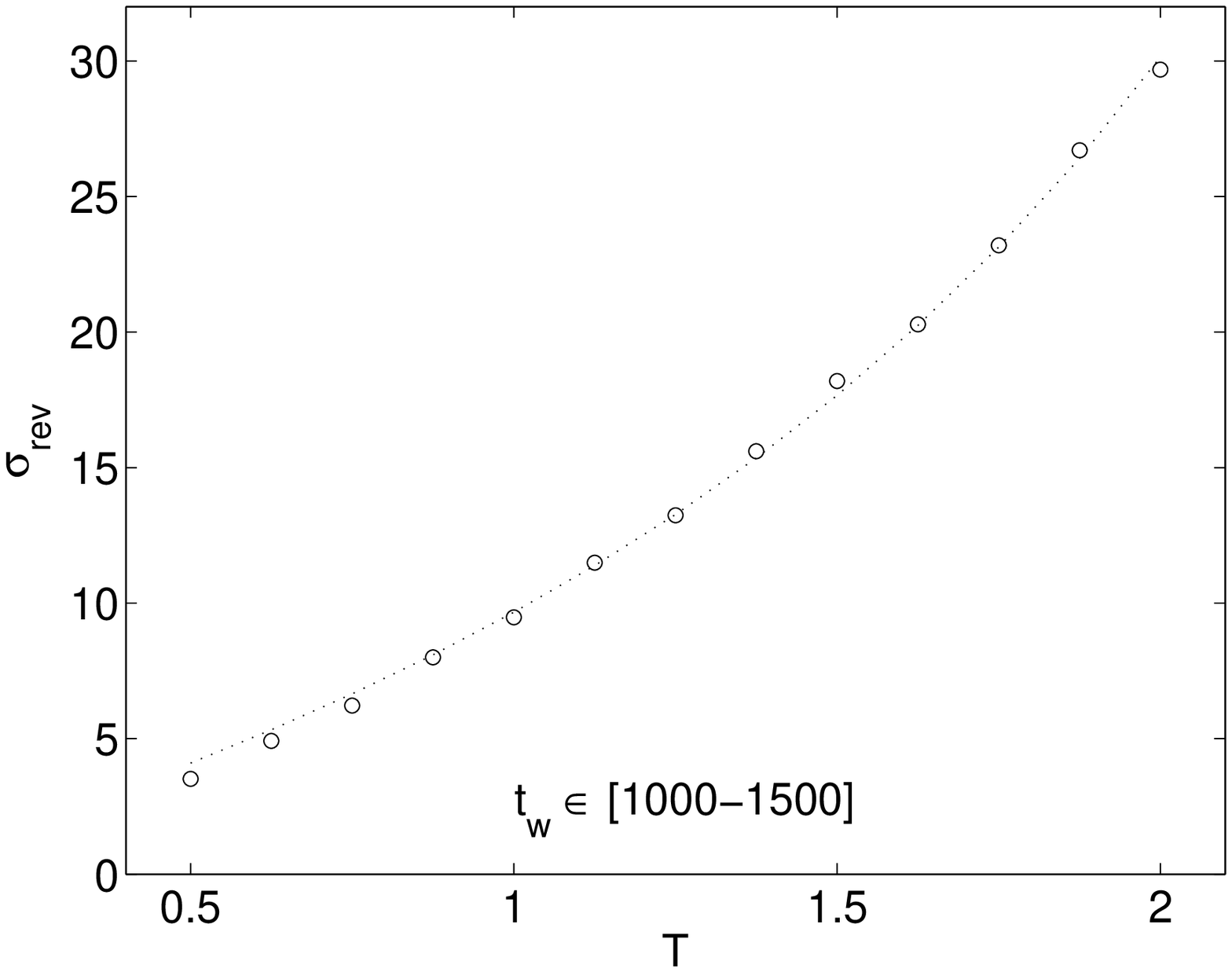,width=.45\textwidth}} \quad 
\subfigure[Activated behavior of $r_E$]{\epsfig{figure=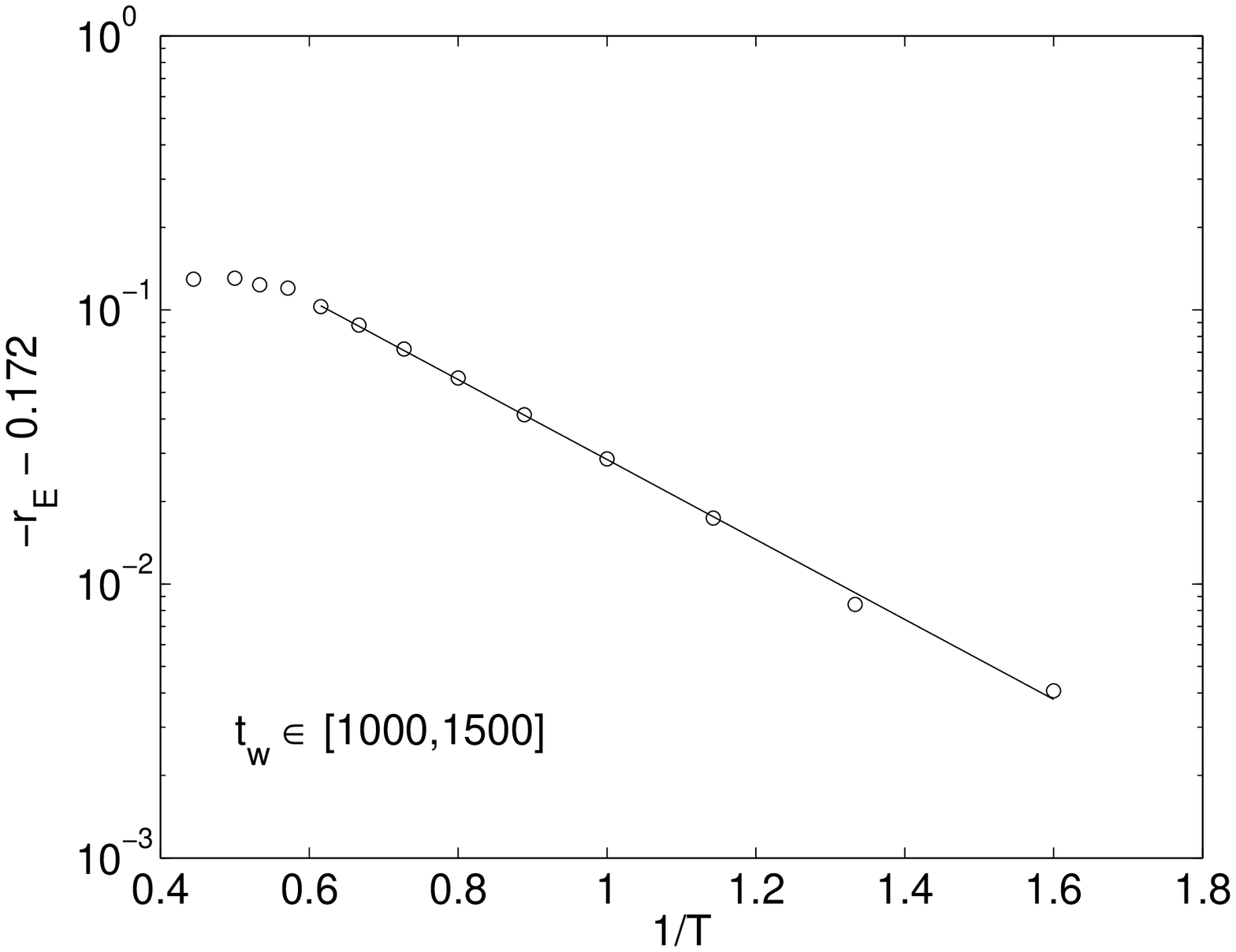,width=.47\textwidth}}} \quad 
\caption{{\em (a)}: The standard deviation $\sigma_{rev}$ of the reversible energy fluctuations
is extracted from the Gaussian fit of the central part of the PDF data, and plotted 
versus the temperature. The dotted line is a fit to the  theoretical prediction
discussed in the main text. The data are based on $800$ independent 
runs,    taken  in the interval   $[1000,1500]$. 
{\em (b)}: The average rate of flow, shifted as indicated,  is 
plotted  logarithmically versus the inverse  temperature. The straight 
line part of the data corresponds to  Arrhenius activated behavior.  
These high precision data are based on $80000$ independent 
runs,  all  taken  in the interval   $[1000,1500]$.  
 }
\label{heat_flow_vs_T}
\end{figure} 
 
Figure~\ref{heat_flow_vs_T} covers 
the temperature dependence of  the   heat transport PDF 
in the age interval $[1000-1500]$. Specifically, panel \emph{(a)}
covers the $T$ dependence of the Gaussian part of the PDF, 
as described by the  corresponding standard deviation
$\sigma_{rev}$, while panel \emph{(b)} covers the    $T$ dependence
of the intermittent tail, as described by the rate $r_E$.

 In panel \emph{(a)} the dots  are the standard deviation $\sigma_{rev}$  of the 
 reversible heat exchange fluctuations
 obtained from the Gaussian part of the PDF. 
The line is the  theoretical prediction 
obtained under   two  model assumptions:  
\emph{(i)}  reversible energy fluctuations are treated as  thermal equilibrium fluctuations,
which  \emph{(ii)}   occur independently within the previously introduced
number $\alpha$ of thermalized 
domains. These are   characterized by an exponential local density of states,
${\cal D}(e) \propto \exp(e/\epsilon)$. These assumptions lead, via standard
thermal equilibrium calculations, to the expression  
\begin{equation}
\sigma_{rev} = \surd (2\alpha) \frac{\epsilon T}{\epsilon -T}; \quad T < \epsilon .
\label{reversible}
\end{equation}  
 The parameter  $\epsilon$   marks   
 the upper   limit of  the temperature range 
 where   the  exponential density  of states is physically 
 relevant (the attractor  is thermally   unstable and 
 physically irrelevant  for $T \ge  \epsilon$). 
The pre-factor $\surd 2$ arises since $H$ is the difference of two
energy values, taken $\delta t$ apart.  These values   are 
statistically independent, and the variance is correspondingly additive, 
 provided that   $\delta t$ is sufficiently large, as is the case   here. 
  Finally the pre-factor $\surd \alpha $ expresses  the linear dependence 
of the variance on the number $\alpha $ of    (supposedly) equivalent 
and independent domains.
The parameter values minimizing the RMS distance between data and Eq.~\ref{reversible}
are  $\alpha=25$  and $\epsilon=3.8$. The    domain 
volume  corresponding to $\alpha=25$ is  $v = 16^3/25 \approx  162$ spins, whence  the linear  
size  is,  for a roughly cubic domain,  close to $5$.  
The $\epsilon$ value is a theoretical  upper bound to the  
 thermal stability range, and safely exceeds  $T=2.5$, the empirical upper temperature limit 
 to the aging regime.
 Finally, using Eq.~\ref{energyperquake}, the value $\alpha \approx 25$ and the 
 available values of $r_E$,   the average energy 
exchanged in a quake, $e(T)$, turns out to    vary from $-8.8$ to $-15$ when $T$ varies from $0.65$  to $2.35$.
In all cases  $-e(T)>>T$, meaning that quakes are  
thermally irreversible on the time scale at which the occur, a necessary 
condition for record dynamics to apply~\cite{Sibani03,Sibani05,Sibani06}. 
 In summary, the  procedure just described   seems 
preferable to  e.g.  a parabolic fit, which has  nearly the same  quality,  but 
 requires one more  parameter, without  providing  a  physical interpretation. 
 
Panel \emph{(b)} of  Fig.~\ref{heat_flow_vs_T} shows the temperature dependence 
of   the  energy flow rate $r_E$ in the age  interval $[1000-1500]$.  
For fixed $t_w$  
and  $T\rightarrow 0$, $r_E$  approaches the    value  $\approx -0.172$. 
The logarithm of the deviation   $-r_E - 0.172$ 
 is plotted in an Arrhenius fashion versus the reciprocal temperature $1/T$ 
 (circles). 
 The full line  in  panel \emph{(b)} is  
 an Arrhenius fit  within the restricted range of $T$
obtained by excluding the lowest temperatures, as indicated. The fit yields
an  energy barrier of characteristic size  $b\approx 3$. The modest  activated contribution  shows that 
  once initiated, a quake can  cover  more ground  
the  higher temperature,  of course within  the boundary  of  the aging regime. 
 The activated contribution   explains   why, 
 within   a fixed amount of time,  aging reaches 
lower energies at higher temperatures (see e.g. Fig. \ref{aging_of_energy}).
In summary, the main  contribution to the energy release is $T$ independent, but
 the process is  nevertheless  enhanced  by 
 an increased ability to climb   small energy barriers.  
 \section{Discussion} 
The  heat exchange  in the plaquette model 
features  both equilibrium-like fluctuations, the Gaussian 
part of the PDF,  and quakes, the exponential tail.  
Crucially,  as the Gaussian  signal has  zero average, 
the energy outflow  is  exclusively due 
to the quakes, which we interpret~\cite{Sibani05,Sibani03}  as a 
change from one metastable configuration to another,
having  considerably lower energy.  
The interpretation is explicitely verified in  a numerical study of 
  the random orthogonal model~\cite{Crisanti04}, where energy changes  from one inherent 
state (IS)~\cite{Stillinger83}  to another
care monitored.  The  processes 
corresponding to the Gaussian and exponential parts of the PDF are 
there  called stimulated and spontaneous, respectively. 

As quakes are  localized events in time and space, 
record dynamics  treats   aging   as a strongly 
heterogeneous dynamical process. 
However, as emphasized by 
recent work~\cite{Mayer04,Appignanesi04,Pan05, Merolle05},
exponential  tails in the PDF's of relevant dynamical  variables 
also appear   outside the aging
regime (see also the insert of Fig.~\ref{aging_of_energy}), i.e. 
heterogeneity is not    exclusively  associated to aging.

Our  analysis  employs a record dynamics scenario~\cite{Sibani03,Sibani05}, 
where quakes are assumed to be  irreversible, an  assumption appropriate
 far from equilibrium, and explicitely verified in the present model.  
 The  scenario has been   used to design a   numerical exploration 
 technique~\cite{Dall03,Boettcher05},  melds, as we have seen,  
 real space (the mumber $\alpha$ of domains) 
 and `landscape' (the density of states of local attractors)  properties of glassy dynamics,  
 and has a number of   predictions 
 which can be tested numerically and/or  experimentally.  
The  physical adjustments  induced  
by each quake affect all  measurements. In principle, these aspects require a  
separate modeling effort, which in our case is limited to 
the  temperature dependence of the average energy $e(T)$ exchanged in a single
quake.  
More generally, by  assuming   that the induced changes are stochastic and 
 mutually   independent, and  that no rearrangement can occur   
 without    a quake,  approximate   closed-form expressions 
 for the rate of change of any  quantity of physical interest
 can be obtained~\cite{Sibani06,Sibani06a}. 
E.g.,  the configuration autocorrelation function
and the linear response between times $t_w$ and $t_w+t$ are predicted to scale with the  
logarithm of the ratio $t/t_w$, as indeed  observed~\cite{Sibani06,Castillo03,Chamon04}
in a number of cases. 
 
In the p-spin   model, the average energy decays in logarithmic fashion,  
a behavior also seen in      spin-glasses~\cite{Sibani05} 
and    Lennard-Jones glasses~\cite{Angelani03}. In the latter case,
the  rate of hopping between different minima basins 
(arguably similar to our quakes, see also ref.\cite{Crisanti04} ) 
 decreases as the reciprocal of the age. 
Broadly similar  behavior is observed, experimentally
or numerically,  in a large number 
of other complex systems~\cite{Lee05,Hannemann05,Oliveira05,Balankin05}.

 Record dynamics sees   aging as an    entrenchment
 into a hierarchy of metastable attractors with growing degree of stability.    
It thus implies  the existence of a hierarchy  
of the sort  explicitely built into  `tree' models
of glassy relaxation~\cite{Sibani89,Hoffmann90,Sibani91,Joh96,Hoffmann97}, 
models which  notably reproduce key features of glassy dynamics. 
For  such  models,  an entry in a hitherto unexplored   subtree  
 requires    a record-sized energy fluctuation. The converse statement,
which is assumed in record dynamics, 
  is  only   true for 
 a continuously branching tree, i.e. when the energy difference 
between  neighboring nodes   tends to zero. In spite of these 
mathematical difference, the physical picture behind record
dynamics and tree models is  nearly the same. Note  
that hierarchies    are specifically associated  
to    locally thermalized   and independently relaxing real space 
domains. 
The fit in Fig.~\ref{heat_flow_vs_T}  simplistically assumes  that,
within each of these, the number of configurations available  
 increases exponentially with the energy difference from the state of lowest
 energy within the domain. The same  property is    built in 
  tree models, and  leads~\cite{Grossmann85} to a
 thermal instability  when the temperature 
approaches from below the characteristic scale of the exponential growth
($\epsilon$ in our notation). 
Nearly exponential local density of states have  been found by a variety of numerical 
techniques  in a  number of microscopic 
models~\cite{Sibani93,Sibani94,Schon98,Klotz98,Schon00,Schon02,Schubert05},
providing independent evidence of the physical relevance of the approximation. 
 
In conclusion,   intermittency data  offer  a unique 
window  into  the microscopic  dynamics of aging 
systems,  and  open the possibility   
of   investigating issues of central 
importance in the   statistical physics  of complex systems,
e.g.  barriers and attractor basins, using calorimetric experiments.
 
\section{Acknowledgments} Financial support  from the Danish Natural Sciences Research Council
is gratefully acknowledged. The author is  greatly indebted  
to Stefan Boettcher for suggesting 
this investigation  and for his useful comments,
and to Karl Heinz Hoffmann, Henrik J. Jensen and  Christian Sch\"{o}n for discussions.
\bibliographystyle{unsrt}
\bibliography{SD-meld}
\end{document}